\documentclass[preprint,showpacs,preprintnumbers,amsmath,amssymb,superscriptaddress]{revtex4}
\usepackage[utf8]{inputenc}
\usepackage[T2A]{fontenc}

\include{ams}
\usepackage{amsmath,amssymb,amsthm}
\usepackage{dcolumn}
\usepackage{bm}
\usepackage{graphicx}

\usepackage{ulem}
\usepackage{color}

\begin{document}
\title{Vacancy-induced enhancement of thermal conductivity in graphene}
\author{S.E. Krasavin}\email{krasavin@theor.jinr.ru}
\affiliation{Bogoliubov Laboratory of Theoretical Physics, Joint
Institute for Nuclear Research, 141980 Dubna, Moscow region, Russia}
\author{V. A. Osipov}\email{osipov@theor.jinr.ru}
\affiliation{Bogoliubov Laboratory of Theoretical Physics, Joint
Institute for Nuclear Research, 141980 Dubna, Moscow region, Russia}

\date{\today}


\begin{abstract}
It is shown that the experimentally observed increase of the Young's modulus in single-layer graphene with low density of point defects leads to a noticeable enhancement of the thermal conductivity in a wide temperature range.
 \end{abstract}

\maketitle

Recently, the effect of mechanical stiffness augmentation in graphene by controlled creation of a low density of point vacancy defects through Ar$^+$ irradiation has been experimentally revealed~\cite{nature}. It has been found that the Young's modulus (E$_{2D}$) of the graphene membrane increases with increasing irradiation dose and reaches a maximum of 550 Nm$^{-1}$ at 0.2\% defect content.
For a higher defect content a decreasing E$_{2D}$ has been observed.
This effect was attributed to suppression of the out-of-plane fluctuations by defects.

The atomic simulation shows that this effect is mainly originated from specific
bonds distribution in the surrounded monovacancy defects ~\cite{Kvashnin}.
Moreover, it has been shown that such  unusual mechanical response is the feature of presence
of specifically monovacancies, whereas other types of point defects such as divacancy,
555-777 and Stone-Wales defects lead to the ordinary degradation of the graphene mechanical stiffness~\cite{Kvashnin}.

Notice that this unusual behavior will affect other important properties of defected graphene.
In this letter, we consider a possible impact of monovacancy defects at a tiny
concentration on the phonon thermal conductivity of graphene. Physically,
a growing number of point defects will enhance the phonon scattering thus leading to the
reduction in the thermal conductivity. In our case, however, the increase of the Young's modulus will
result in the increased sound velocities, which reduces anharmonic phonon-phonon scattering processes and, thereafter, enhances the thermal conductivity. These opposite effects
will compete in a wide temperature range. The aim of our paper is to analyze the influence of point
defects on the thermal conductivity of graphene within a phenomenological single-mode relaxation time approach with all important scattering mechanisms taken into account.

Let us start with the well-known definition of sound velocities for longitudinal $(LA)$ and transverse $(TA)$ phonon branches in isotropic case~\cite{munoz}
\begin{equation}
v_{LA}=\sqrt{E_{2D}/\rho (1-\sigma^2)}, \quad
v_{TA}=\sqrt{E_{2D}/2\rho (1+\sigma )},
\end{equation}
where $\sigma $ is the Poisson's constant, $\rho $ is the material density and the Young's modulus is assumed to be a function of the defect density $n_{def}$. The inset in Fig. 1 shows the fit to the measured $E_{2D}$ as a function of defect concentration given in~\cite{nature}.

We use the Callaway's theory where three phonon normal processes are taken into consideration explicitly~\cite{call}. Notice that the important role of normal phonon scattering processes in graphite-like materials and graphene has been noted in~\cite{slack,morelli,alofi}. Thus, we consider four main scattering mechanisms relevant for suspended graphene: sample border (rough boundary), point defects, three-phonon normal and umklapp processes. Within the relaxation-time approximation the total mean free path can be written as
\begin{equation}
l_{tot,\lambda }^{-1}(q)= l_{0}^{-1}+l_{pd,\lambda }^{-1}(q)+l_{N,\lambda }^{-1}(q)+l_{U,\lambda }^{-1}(q),
\end{equation}
where $l_{0}$, $l_{pd,\lambda }$, $l_{N, \lambda }$ and $l_{U, \lambda }$ come from sample border, point defects, three phonon normal and umklapp  scattering, respectively, for a given phonon branch $\lambda =(LA,TA,ZA)$ with the wave vector $q$.
The mean free path due to normal  processes is written as (see, e.g.,~\cite{morelli, alofi})
\begin{equation}
l_{N,\lambda }^{-1}(q)=B_{N,\lambda }\omega^{2}_{\lambda}(q)T^{3},
\end{equation}
and for umklapp phonon scattering processes we employ a parametrized expression in the form
\begin{equation}
l_{U,\lambda }^{-1}(q)=B_{U,\lambda }T^{3}\omega ^{2}_{\lambda}(q)\exp(-\Theta _{\lambda }/3T),
\end{equation}
where $B_{N,\lambda }$ and $B_{U,\lambda }$ are parameters and $\Theta _{\lambda }$ is the Debye temperature.
It should be mentioned that in our case $B_{(N,U)\lambda }=\bar{B}_{(N,U)\lambda }/v_{\lambda}$, and the numerical values of the parameters $\bar{B}_{N,\lambda }$ and $\bar{B}_{U,\lambda }$ are taken from~\cite{alofi}: $\bar{B}_{N,\lambda}=2.12\times 10^{-25}s$K$^{-3}$, $\bar{B}_{U,\lambda }=3.18\times 10^{-25}s$K$^{-3}$ for $\lambda =(LA,TA)$ and $\bar{B}_{N,\lambda}=1.48\times 10^{-22}s$K$^{-3}$, $\bar{B}_{U,\lambda }=2.23\times 10^{-22}s$K$^{-3}$ for $\lambda =(ZA)$.

The boundary scattering is expressed as
\begin{equation}
l_{0}^{-1}=\frac{1}{d}
\end{equation}
with $d$ being the effective length determined from the geometry of the graphene sample~\cite{alofi}. The mean free path due to phonon-point defect scattering is taken to be
\begin{equation}
l_{pd,\lambda }^{-1}(q)=\frac{S _{0}\Gamma }{4}\frac{q\omega _{\lambda }^{2}(q)}{v_{\lambda }^{2}},
\end{equation}
where $S _{0}$ is the cross-section area per one atom of the lattice, $\Gamma\approx (n_{def}/2)\times 10^{-15}$cm$^{-2}$ is the mass-fluctuation phonon-scattering parameter (notice that $1 \%$ of vacancies corresponds to $n_{def}=2\times 10^{13}$cm$^{-2}$~\cite{morelli}).

Within Callaway's formalism, the diagonal components of the thermal conductivity tensor $\kappa(T)$ can be presented by the sum of the Debye term
\begin{equation}
\kappa _{D}(T) = \frac{\hbar ^2}{S_{0}k_{B}T^{2}}
\sum_{\lambda }\int d\omega \omega _{\lambda }^{2}(q)
 l_{tot,\lambda }(\omega )v_{\lambda }C_{ph,\lambda }(\omega )N _{\lambda }(\omega ),
\end{equation}
and the normal-drift term
\begin{equation}
\kappa _{N}(T) = \frac{\hbar ^2}{S_{0}k_{B}T^{2}}
\sum_{\lambda }\frac{\Bigl[\int d\omega \omega _{\lambda }^{2}(q)
 l_{tot,\lambda }(\omega )l_{N,\lambda }^{-1}(\omega )v_{\lambda }^{2}C_{ph,\lambda }(\omega )N _{\lambda }(\omega )\Bigr]^{2}}{\int d\omega \omega _{\lambda }^{2}(q)
 (1-l_{tot,\lambda }(\omega )l_{N,\lambda }^{-1}(\omega ))l_{N,\lambda }^{-1}(\omega )v_{\lambda }^{3}C_{ph,\lambda }(\omega )N _{\lambda }(\omega )},
\end{equation}
where $C_{ph,\lambda }(\omega )=\exp(\hbar \omega _{\lambda}(q)/k_{B}T)/(\exp(\hbar \omega _{\lambda}(q)/k_{B}T)-1)^2$, $N_{\lambda }(\omega )$ is the density of states function per mole for each phonon branch, $k_{B}$ is the Boltzmann constant. Summation is performed over phonon polarization branches with the dispersion relations $\omega _{\lambda }(q)=qv_{\lambda }$  for $\lambda=LA,TA $. For out-of-plane (flexural) acoustic mode we use the dispersion law $\omega _{ZA }(q)=q^2/2m$~\cite{nic} ($m$ is an effective parameter taken here  equal to $320$ sec/cm), $l_{tot,\lambda }(q,T)$ is the phonon mean free path given by Eqs.(2)-(6). The explicit form of $N_{\lambda }(\omega )$ is taken from~\cite{alofi}.


Fig. 1 shows the calculated $\kappa (T)$ based on Eqs. (7) and (8) at the fixed concentration of vacancies $n_{def}=1.5\times 10^{13}$ cm$^{-2}$ for two cases: (a) the sound velocities do not depend on $n_{def}$ and have fixed values taken from~\cite{nika1}: $v_{LA}=21.3\times 10^{5}$ cm/s, $v_{TA}=13.6\times 10^{5}$ cm/s  (which corresponds to the Young's modulus $E_{2D}\approx 360$ Nm$^{-1}$), and (b) the sound velocities are calculated by  Eq.(1).
\begin{figure} [tbh]

\begin{center}
\includegraphics [width=11 cm]{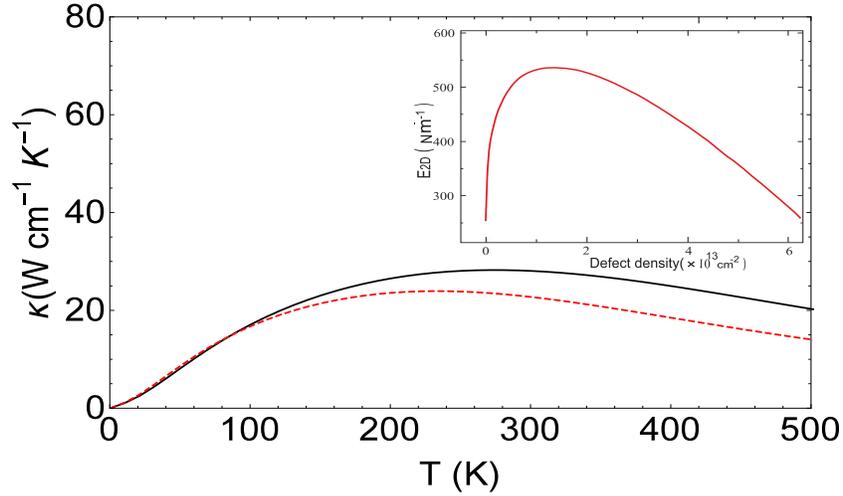}
\end{center}
\caption{Thermal conductivity versus temperature in a 2.9 $\mu$m wide ribbon at $n_{def}=1.5\times 10^{13}$ cm$^{-2}$ in case of actual (solid line) and constant (dashed line) sound velocities.
The insert shows a fit to the experimentally observed Young's modulus as a function of defect concentration (cf. Ref.~\cite{nature}).}
\end{figure}
For chosen $n_{def}$ one has $E_{2D}\approx 540$ Nm$^{-1}$, so that $v_{LA}=27.2\times 10^{5}$ cm/s and $v_{TA}=17.2\times 10^{5}$ cm/s. As seen in Fig.1, in case of (b) markedly enhanced thermal conductivity takes place in a wide temperature range. The reason is quite clear because larger values of $v_{LA}$ and $v_{TA}$ lead to an increase of $l_{N,\lambda }(q)$ and $l_{U,\lambda }(q)$. At higher concentrations of vacancies, the difference between two cases (a) and (b) disappears  which agrees with experimentally observed behavior of the Young's modulus (see the insert in Fig.1).

Fig.2 shows the thermal conductivity as a function of $n_{def}$ at $T=300$K.
\begin{figure} [tbh]
\begin{center}
\includegraphics [width=11 cm]{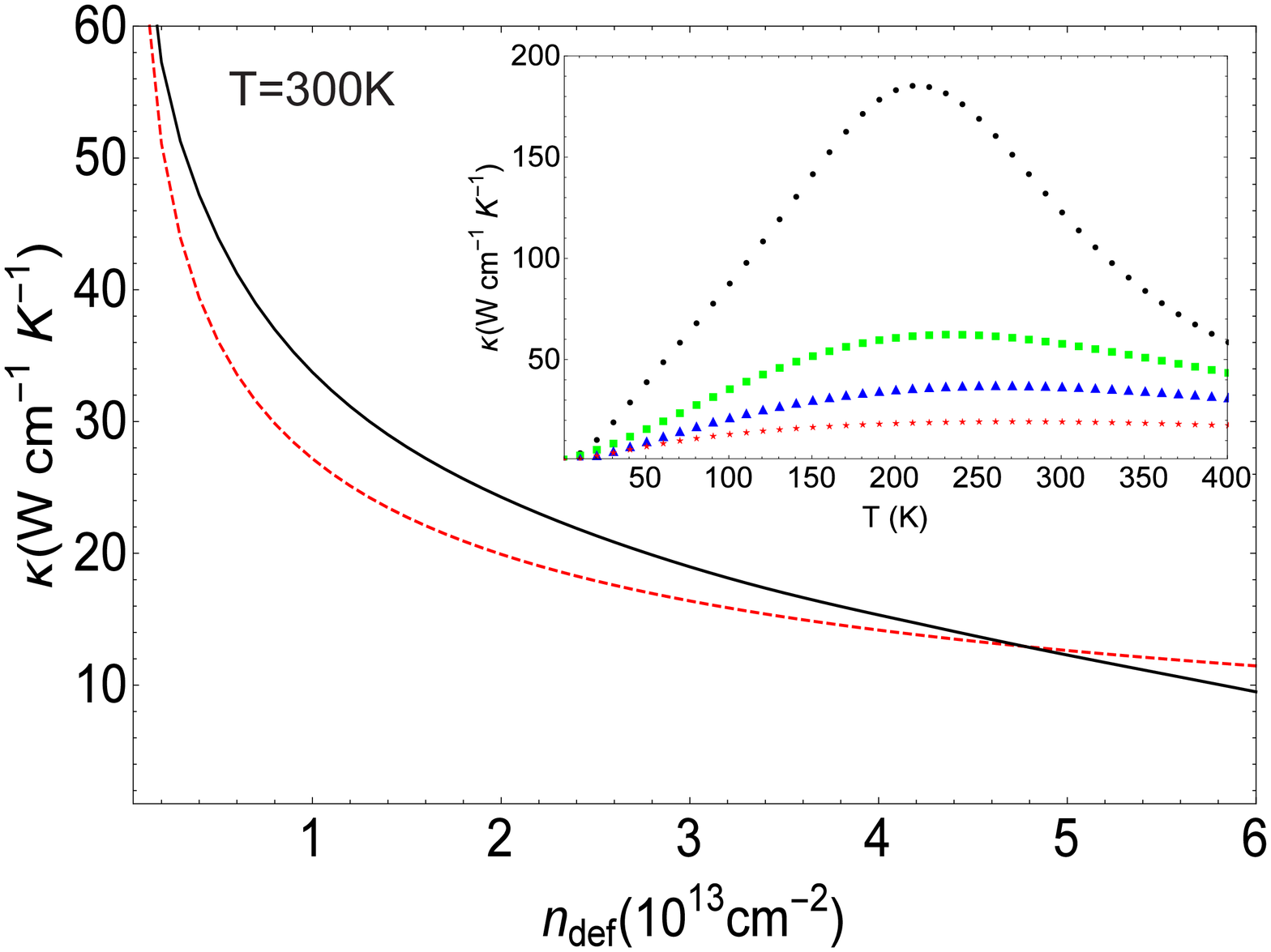}
\end{center}
\caption{Thermal conductivity versus the concentration of vacancies in a 2.9 $\mu$m wide ribbon for actual (solid line) and constant (dashed line) sound velocities. The insert shows the calculated thermal conductivity versus temperature with the experimentally observed values of the Young's modulus for $n_{def}=0$ (circles), $2\times 10^{12}$ cm$^{-2}$ (squares), $8\times 10^{12}$ cm$^{-2}$ (triangles), $3\times 10^{13}$ cm$^{-2}$ (stars).
}
\end{figure}
As is seen, the enhancement of $\kappa $ takes place in the range of $2.8\times 10^{12}$ cm$^{-2}$$\le n_{def}$$\le 5.3\times 10^{13}$ cm$^{-2}$ only. The explanation is as follows: at high temperatures the normal and umklapp scattering mechanisms are of the most importance and, in this case, $\kappa $ strongly depend on  $v_{\lambda}$ (see, e.g.,~\cite{nika1}). In turn, $v_{\lambda }$ is a function of $E _{2D}$ in accordance with Eq. (1) and therefore depends on $n_{def}$ in this region. Accordingly, we have obtained the increased  $\kappa $ at small concentrations of vacancies up to the value close to $5\times 10^{13}$ cm$^{-2}$ with the maximum gain at $n_{def}\sim 7\times 10^{12}$ cm$^{-2}$ which corresponds to the maximum of the Young's modulus.

Summarizing, we have found a marked increase in the thermal conductivity of graphene, which is a direct consequence of the experimentally observed effect of ultrahigh stiffness at low densities of vacancy defects. Our study shows that, in a limited range of defect concentrations, the thermal transport demonstrates a rather unique behavior. Namely, the growing number of defects provokes the enhancement of the thermal conductivity in a wide temperature range. Physically, this follows from the fact that after about $100$ K the three-phonon scattering processes  become dominant. They depend on the sound velocities which grow with the Young's modulus in some restricted region of $n_{def}$. This provides the enhancement of the thermal conductivity. Below $T\le 100$ K the main sources of the phonon scattering are sample border and point defects so that a strong increase in graphene stiffness has no effect on the thermal conductivity. Notice that our finding can be of importance in development of graphene-based thermoelectric devices.



%
%

 \end{document}